\documentclass{aa}
\usepackage{graphicx}

\usepackage[varg]{txfonts}

\newcommand{\Zs}{\ensuremath{Z_{\odot}}}

\newcommand{\TrOiii}{$T_{\rm{r}}$[O~{\sc iii}]}

\newcommand{\Hb}{\ifmmode {\rm H}\beta \else H$\beta$\fi}
\newcommand{\hi}{\ion{H}{i}}
\newcommand{\hii}{\ion{H}{ii}}

\newcommand{\Oii}{[\ion{O}{ii}] $\lambda$3727}

\newcommand{\Oiii}{[\ion{O}{iii}] $\lambda$5007}
\newcommand{\Oiiit}{[\ion{O}{iii}] $\lambda$4363}

\newcommand{\ariv}{[\ion{Ar}{iv}]}

\newcommand{\rOiii}{[\ion{O}{iii}] $\lambda$4363/5007}
\newcommand{\rNii}{[\ion{N}{ii}] $\lambda$5755/6584}

\newcommand{\Op}{\ensuremath{\mathrm{O}^{+}}}
\newcommand{\Opp}{\ensuremath{\mathrm{O}^{++}}}

\newcommand\HII{H\,{\scshape\bfseries ii}}
\newcommand\ISM{_{\rm ISM}}

\begin{document}

\title{Enrichment of the ISM by metal-rich droplets and 
the abundance bias in \HII{} regions}

\author{Gra\.zyna Stasi\'nska\inst{1}
   \and
   Guillermo Tenorio-Tagle\inst{2}
   \and M\'onica Rodr\'\i guez\inst{2}
   \and William J. Henney\inst{3}}

\offprints{grazyna.stasinska@obspm.fr}

\institute{LUTH, Observatoire de Paris, CNRS, Universit\'e Paris Diderot; Place Jules Janssen 92190 Meudon, France 
\and
Instituto Nacional de Astrof\'\i sica \'Optica y 
Electr\'onica, AP 51, 72000, Puebla, Mexico
\and
Centro de Radioastronom\'{\i}a y 
Astrof\'{\i}sica, Universidad Nacional Aut\'onoma
de M\'exico, Campus Morelia, Apartado Postal 3-72, 58090 Morelia, Mexico}

\date{Received ???; accepted ???}
\titlerunning{ }
\authorrunning{ }


\abstract{ We critically examine a scenario for the enrichment
  of the interstellar medium (ISM) in which supernova ejecta follow a
  long ($10^8$ yr) journey before falling back onto the galactic
    disk in the form of metal-rich ``droplets'', These droplets
    do not become fully mixed with the interstellar medium until they
    become photoionized in \hii{} regions.  We investigate the
    hypothesis that the photoionization of these highly
  metallic droplets can explain the observed ``abundance
    discrepancy factors'' (ADFs), which are found when
    comparing abundances derived from recombination lines and from
  collisionally excited lines, both in Galactic and
  extragalactic \hii{} regions. We derive bounds of
    $10^{13}$--$10^{15}$~cm on the droplet sizes inside \hii{} regions
    in order that (1)~they should not have already been detected by
    direct imaging of nearby nebulae, and (2)~they should not be too
    swiftly destroyed by diffusion in the ionized gas. From
    photoionization modelling we find that, if this
    inhomogeneous enrichment scenario holds, then the
  recombination lines strongly overestimate the metallicities of the
  fully mixed \hii{} regions.  The abundances derived from
  collisionally excited lines also suffer some bias, although to a
  much lesser extent. In the absence of any recipe for correcting
  these biases, we recommend the discarding of all objects showing
  large ADFs from studies of galactic chemical evolution. These biases
  must also be kept in mind when comparing the galactic abundance
  gradients for elements derived from recombination lines with those
  derived from collisionally excited lines.  Finally, we propose
  a set of observations that could be undertaken to test our
  scenario and improve our understanding of element mixing in the ISM.

}
\maketitle

\section{Introduction}
\label{sec:introduction}
The detailed process of enrichment of the 
interstellar medium (ISM) by products arising 
from supernovae explosions and stellar winds is 
far from being fully understood (see review by 
Scalo \& Elmegreen 2004). Many mechanisms are 
likely to be at work on different scales, both in 
time and in space (see, e.g., Bateman \& Larson 
1993; Roy \& Kunth 1995; Tenorio-Tagle 1996; de 
Avillez \& Mac Low 2002).

The present-day chemical composition of the ISM 
is derived from the analysis of the chemical 
abundances of \hii{} regions. In such analysis, 
it is always assumed that \hii{} regions are 
chemically homogeneous. So far, 2D and 3D 
spectroscopy of \hii{} regions has not revealed 
the existence of zones with significantly 
different abundances within the same \hii{} 
region, except in the case of NGC 5253, which 
shows some local N enhancement (Walsh \& Roy 
1989; Kobulnicky et al.\@ 1997, L\'opez-S\'anchez 
et al.\@  2007). On the other hand, chemical 
inhomogeneities occuring on scales smaller than 
the spatial resolution would pass unnoticed when 
comparing spectra of neighbouring zones.

There is, however, a way to unravel the presence 
of small-scale abundance inhomogeneities in 
\hii{} regions. This is by comparing the 
abundances derived by traditional methods, i.e. 
from collisionally excited lines (CELs), to those 
obtained from optical recombination lines (ORLs) 
in the same spectrum. CELs are preferentially 
emitted in zones of high electron temperature and 
low metallicity, while ORLs are preferentially 
formed in zones of low electron temperature and 
high metallicity. So far, ORLs have been observed 
in only a few \hii{} regions. They systematically 
lead to an abundance discrepancy factor 
(ADF)\footnote {The abundance disprepancy 
factor (ADF; Tsamis et~al.\@ 2004) of an ion 
is defined as the ratio of its abundance as 
derived from the intensities of ORLs to that 
derived from CELs. } larger than one. This has 
first been interpreted as due to temperature 
fluctuations within the \hii{} regions (Peimbert 
et~al.\@ 1993; Peimbert 2003; Esteban et al.\@ 
1998, 1999a, 1999b, 2002, 2004). However, Tsamis 
et~al.\ (2003) consider that temperature 
fluctuations alone cannot be the cause of the 
observed discrepancies, since in that case oxygen 
abundances derived from far-infrared lines should 
be close to the high values obtained from optical 
recombination lines, and this is not the case 
(however, aperture problems are difficult to deal 
with, see Garc\'\i a-Rojas et~al.\@ 2006). Tsamis 
et~al.\ argue for the existence of a hitherto 
unseen component in \hii{} regions, consisting of 
cold, metal-rich ionized parcels of gas. In fact, 
chemical inhomogeneities have been proposed for 
over a decade to explain the ORL/CEL discrepancy 
in planetary nebulae (Torres-Peimbert et al.\ 
1990; Liu et al.\ 2000, 2004), but such an 
interpretation in the case of \hii{} regions has 
appeared only recently (Tsamis et al.\@  2003; 
Tsamis \& P\'equignot 2005).
Of course, chemical inhomogeneities in \hii{} 
regions and planetary nebulae must have a very 
different origin.
 Note that invoking chemical inhomogeneities accounts at the same time 
 for the origin of temperature fluctuations, which otherwise are 
 difficult to explain quantitatively (see Stasi\'nska 2007 and references therein).

 In this article, we explore the possibility that the ADFs measured in
 \hii{} regions result, as suggested by Tsamis et al. (2003, 2005),
 from the presence of metal-rich ``droplets''\footnote{The word
   ``droplets'' is used in this paper with the same meaning as in
   Tenorio-Tagle (1996). It does not imply that they are in liquid
   form, but merely that they are of lower temperature and higher
     density than their surroundings.} as predicted in the scenario
 of Tenorio-Tagle (1996; hereafter referred to as T-T96; see also
 Stasi\'nska et al. 2007) for the enrichment of the ISM by Type II
 supernovae. Note that, up to now, reliable abundance discrepancy
 factors in \hii{} regions have been measured only for oxygen. For
 simplicity, we will then consider only oxygen in this paper.
 Section~2 reviews the framework that accounts for the origin of such
 inhomogeneities in \hii{} regions and evaluates their survival
 time-scale until full mixing with the ISM. This section gives also
 some observational arguments in support of the model. Section 3
 estimates how much matter can be expected in the form of metal-rich
 droplets. Section 4 discusses the ORL/CEL discrepancy in \hii{}
 regions in the context of metal-rich droplets.  Section 5 presents
 some final comments and prospects.

\section{Metal-rich droplets in \HII{} regions}
\label{sec:metal-rich-droplets}
\subsection{Formation of the droplets}
\label{sec:formation-droplets}

The scenario proposed by T-T96 accounts for the fact that the ejecta
from type II supernovae (SNe) ought to follow a long excursion into
the galactic environment before they are able to mix with the ISM.
The excursion is promoted by the clustering of massive stars and thus
of type II SNe, which, being correlated in space and time, force the
violently ejected matter to generate large-scale superbubbles. These
are able to exceed the thickness of galactic disks and burst into the
haloes of the host galaxies, while displacing and locking the
surrounding ISM into kpc-size expanding supershells.  Such remnants,
driven by the hot superbubble interior grow for as long as massive
stars continue to release their metals, until the last 8~$M_\odot$
star belonging to the star cluster or OB association completes its
evolution ($\sim$ 40 Myr).  During this time, the metals injected into
superbubbles have enough time to mix with the matter from stellar
winds as well as with the matter thermally evaporated into the
superbubbles from their surrounding outer supershells. Mixing within
the superbubble interior is strongly favored by the high temperatures
($T \sim 10^6$--$10^7$ K) and high sound speeds and by the stirring
caused by the bursting of more SNe.  However, the remaining ISM, being
out of contact with the superbubble interior, is left uncontaminated
by the products of the evolving star cluster.  After the last SN
explosion, the low density within the superbubble interior ($n \sim
10^{-2}$--$10^{-3}$ cm$^{-3}$) delays the impact of radiative cooling
($t_{\Lambda} \sim k T/ (\Lambda n) \sim 10^{8}$ yr; where $\Lambda$
is the cooling rate and $k$ the Boltzmann constant). Note also that
radiative cooling does not occur at the same rate within the entire
volume of the superbubble, but rather zones with a higher density will
cool faster than their surroundings. Such a situation has been shown
to lead, in other astronomical circumstances such as galaxy formation
or globular cluster formation (see Zel'dovich \& Raizer 1966, Vietri
\& Pesci 1995), to multiple re-pressurising shocks which in our
scenario would lead to the formation of small, dense
``cloudlets'' of metal-rich gas, in pressure equilibrium with
the remaining hot gas. The high density of the cloudlets causes
them to recombine, further reducing their temperature. The cold
cloudlets would inevitably begin to fall towards the disk of the
galaxy under the action of gravity. As they fall through the lighter
interstellar medium, they will be subject to various instabilities,
such as Rayleigh-Taylor and Kelvin-Helmholtz (e.g., R\'o\.zyczka \&
Tenorio-Tagle 1985; Schiano et al. 1995), which will lead to
their breakup into a swarm of minute droplets.

The overall outcome of such a fountain with a 
spray (as described by Scalo \& Elmergreen 2004) 
is a large-scale dispersal of the matter 
processed by the star cluster over a large 
(kpc-scale) galactic volume, a volume much larger 
than the size of the stellar association.  This 
is very different from other fountain models in 
which the hot gas rises to great altitudes 
($\sim$ 3 kpc) above the galactic plane to then 
cool all at once (see Kahn 1991). Such fountains 
would lead to massive giant clouds falling 
unimpeded towards the disk of galaxies. Note also 
that none of those models have addressed the 
mixing of metals with the ISM. On the other hand, 
the fountain with a spray of \mbox{T-T96}, 
predicts a stratus of dense, highly metallic and 
perhaps molecular droplets interspersed in the 
galactic ISM, wherever it had rained. Once this 
happens, the droplets are to participate in the 
general motion of clouds and intercloud medium 
promoted by galactic rotation and stellar 
activity, as envisaged by Roy \& Kunth (1995). 
This is to lead to an even larger dispersal, 
although not to mixing of the metals with the 
ISM. For total mixing a new episode of stellar 
formation seems required as this will cause a 
thorough dissemination of the metals into the ISM 
enhancing its metallicity. This is primarily due 
to the UV radiation field, able to disrupt 
molecules and ionize the droplets and their 
surroundings, while generating multiple localized 
champagne flows (Tenorio-Tagle 1979) and with 
them the stirring to promote a rapid mixing 
within the ionized volume.

\subsection{Survival of the droplets in \hii{} regions}
\label{sec:surv-dropl-hii}

When the droplets are ionized, they will quickly (on order of a
  sound-crossing time) reestablish pressure balance with their
  surroundings.  The two processes that will tend to disperse the
high concentration of metals in the droplets are molecular diffusion
and turbulent mixing.

The strong gradient in the concentration of 
oxygen ions between the droplets and the 
surrounding gas will eventually be diffused away 
due to the collisional random walks of the 
particles.  The time required for diffusion 
through a distance $L$ is of order
\begin{equation}
   \label{eq:diffusion-time}
   t_\mathrm{d} = L^2 / D ,
\end{equation}
where $D$ is the diffusion coefficient, which can 
be considered as the product of the 
root-mean-square particle velocity and the 
mean-free path between collisions (e.g., T-T96). 
For the diffusion of O$^{++}$ ions moving in a 
field of H$^+$ ions, $D$ can be calculated from 
the formulae presented in Oey (2003) as
\begin{equation}
   \label{eq:diffusion-coefficient}
   D = 1.04 \times 10^{17} T_4^{5/2} n^{-1} f^{-1} \mathrm{\ cm^2\ s^{-1}} ,
\end{equation}
where $T_4$ is the gas temperature in units of 
$10^4$~K, $n$ is the hydrogen ion density, and $f 
= 1 + 0.029 \log T_4 - 0.010 \log n$. The 
resultant diffusion time is
\begin{equation}
   \label{eq:diffusion-time2}
   t_\mathrm{d} = 3.05\times 10^5 L_{15}^2 n f T_4^{-5/2} \mathrm{\ years},
\end{equation}
where $L_{15}$ is the length scale in units of $10^{15}\mathrm{\
  cm}$. Adopting characteristic values of density and temperature for
O-rich droplets in pressure equilibrium ($n = 1000 \mathrm{\
    cm^{-3}}$, $T_4 = 0.6$, see Section~4), we find $t_\mathrm{d} =
10^9 L_{15}^2$~years. Therefore, metal-rich droplets as small as
$10^{14}\mathrm{\ cm}$ can survive over the entire lifetime of a
typical \hii{} region, but smaller droplets would be erased on smaller
timescales and therefore would not contribute to the spectrum of any
but the youngest \hii{} regions. The corresponding mass of each
  droplet would be $\sim 10^{-11}~M_\odot$. However, it is possible
that disordered magnetic fields in the ionized gas might increase the
diffusion time still further and allow smaller droplets to survive.

Discounting this last proviso, we have a rather 
stringent condition on the properties of the 
droplets: a physical size of $10^{14}\mathrm{\ 
cm}$ corresponds to an angular size of $0.015''$ 
at the distance of the Orion nebula. Hence, 
droplets that are only a few times larger than 
this  size may in principle be resolvable 
via direct imaging with the \textit{Hubble Space 
Telescope}. There is already some observational 
evidence that inhomogeneities may indeed be 
present at this scale (O'Dell et~al.\@ 2003)

\newcommand\turb{_\mathrm{turb}}
\newcommand\vturb{\ensuremath{v\turb}}
\newcommand\lturb{\ensuremath{\ell\turb}}

Diffusion by turbulent motions can be treated in 
a similar way. Assuming turbulent eddies of 
velocity \vturb{} and size \lturb{}, then for 
times less than the coherence or turnover time, 
$\lturb/\vturb$, matter is simply advected at a 
speed $\vturb$ and so the effective diffusion 
coefficient grows linearly with time. For $t > 
\lturb/\vturb$, on the other hand, the diffusion 
coefficient saturates at a constant value:
\begin{equation}
   \label{eq:turb-diffusion-coefficient}
   D\turb = \lturb \vturb .
\end{equation}
However, this equation only gives the rate of 
\emph{dispersal} of the metal-rich gas by the 
turbulent eddies (Klessen \& Lin 2003), but does 
not directly address the physical \emph{mixing} 
of the two phases (de Avillez \& Mac Low 2002). 
Mere dispersal of the droplets is not enough to 
destroy their physical integrity and distinctive 
temperature. Instead, it is necessary for the 
turbulent eddies to stretch and distort the 
droplets, so as to steepen the concentration 
gradients until they reach a level at which 
molecular diffusion can occur (see above). The relevant timescale for this process can be derived from the
exponential stretching of fluid elements by the turbulent shear (Pan
\& Scalo 2007) as
\begin{equation}
t_\mathrm{mix}\sim t_\mathrm{d} \ln \mathcal{P} .
\end{equation}
In this equation, $t_\mathrm{d}$ is the turbulent diffusion timescale
calculated using equation~(\ref{eq:turb-diffusion-coefficient}) with
$\lturb = L$ (only those turbulent motions on the same scale as the
droplets can efficiently distort them) and $\mathcal{P} =
D_\mathrm{turb}/D$ is the P\'eclet number of the turbulence, with $D$
given by equation~(\ref{eq:diffusion-coefficient}).
If turbulence of velocity $v'$ is driven at some 
large scale $\ell'$, then the turbulent motions 
at a smaller scale $\lturb$ have velocity $\vturb 
= v' (\lturb/\ell')^a$, where $a = 1/3$ for 
incompressible Kolmogorov turbulence (applicable 
to subsonic flows) and $a = 1/2$ in the limit of 
highly supersonic turbulence. The turbulence in 
\hii{} regions is likely driven by the mutual 
interactions of multiple transonic photoablation 
flows (Mellema et~al.\@ 2006; Henney 2006), with 
Mach numbers in the range 2--4 and at scales of 
0.1 to 1~parsec. We therefore find a turbulent 
mixing time of
\begin{equation}
   \label{eq:turb-diffusion-time}
   t_\mathrm{mix} = 12,600 (L_{15} 
\ell'_{18})^{1/2} v_{10}^{-1} g \mathrm{\ years} ,
\end{equation}
where $\ell'_{18}$ and $v_{10}$ are the scale and 
velocity of the turbulent driving in units of 
$10^{18}\mathrm{\ cm}$ and $10\mathrm{\ km\ 
s^{-1}}$, respectively, and $a=1/2$ is 
assumed.\footnote{The expression also includes 
the  term $g = 1 + 0.034 \log n_3 f + 0.052 \log 
L_{15} - 0.017 \log\ell'_{18} - 0.086 \log T_4 $, 
but this will always be close to unity}. This 
result implies that turbulent mixing is efficient 
on timescales shorter than those characteristic 
of \hii{} region evolution.

Since the turbulent $t_\mathrm{d}$ decreases as 
the droplet size is reduced, the turbulence will 
quickly fragment the droplets to such an extent 
that the molecular diffusion also becomes 
efficient (at scales of order $10^{12}\mathrm{\ 
cm}$), at which point chemical mixing will be 
complete. The long-term survival of the droplets 
in the \hii{} region therefore depends critically 
on the avoidance of turbulence.

The extent to which turbulence is present in 
\hii{} regions is somewhat controversial. It is 
certainly the case that the internal kinematics 
of \hii{} regions are highly irregular, with 
significantly non-thermal linewidths (O'Dell \& 
Casta\~neda 1987). However, it is unclear whether 
this represents true hydrodynamical turbulence, 
or simply the superposition of multiple velocity 
components, each one of which arises in an 
ordered flow. Recent simulations of \hii{} region 
evolution in a clumpy molecular cloud (Mellema 
et~al.\@ 2006) indicate that the ``chaotic'' 
velocity field is largely the result of 
inward-pointing photoablation flows, which form 
wherever the outward propagation of the 
ionization front is detained by a dense neutral 
condensation. These flows shock against one 
another in the interior of the \hii{} region, 
which does become truly turbulent. However, it is 
found that 10--30\% of the [\ion{O}{iii}] 
emission comes from the non-turbulent flows 
closer to the ionization front. A similar result 
was derived analytically by Henney (2003), who 
showed that the surface brightness of such 
photoablation flows is proportional to the linear 
size of the cavities that they carve out of the 
\hii{} region, which in turn is a function of the 
scale of the irregularities in the front. Flows 
from condensations with size $> 10\%$ the radius 
of the \hii{} region will have a brightness 
comparable to that of the entire region. Both 
these studies considered only \hii{} regions with 
globally closed geometries, although it is likely 
that many optically visible regions are instead 
globally open champagne flows (e.g., 
Tenorio-Tagle 1979; Henney et~al.\@ 2005), which 
would tend to increase the relative contribution 
of the photoevaporation flows to the total 
emission. On the other hand, the interaction of 
the stellar wind from the ionizing star with the 
other flows (e.g., Garc{\'{\i}}a-Arredondo 
et~al.\@ 2001) may be an additional source of 
turbulence, as may instabilities of the 
ionization front (Garc{\'{\i}}a-Segura \& Franco 
1996; Williams 1999, 2002). Here we thus assume that turbulence and molecular diffusion act very
effectively within the HII region volume, causing a thorough mixing and homogeinizing of the ionized gas. 
However, we also assume that at all times there are new metal-rich droplets traversing the ionization front into 
the HII region.

\subsection{Observational considerations}
\label{sec:observ-cons}
There are several further observational facts that lend support
  to the droplet scenario, facts that ought to be taken into
consideration by other possible explanations of how the mixing of
heavy elements with the ISM takes place. Regarding abundances in
\hii{} regions the issues are:

Given the lifetime  of type II SN progenitors (up 
to several  $10^7$ yr), it is obvious that the 
metals expected from  recent   bursts of stellar 
formation (age $\sim$ a few $ 10^6$ yr) have not 
yet been released and thus the detected 
abundances in \hii{} regions reflect the 
production, dispersal and mixing of heavy 
elements  ejected by former stellar generations.

This result is supported by the work of Herrero 
\& Najarro (2005) and Sim\'on-D\'\i az et~al.\@ 
(2006) who measured  the metallicity of the 
ionizing stars (in Orion and in the largest 
\hii{} regions in M33)   and found that these are 
similar to those determined for their 
corresponding \hii{} regions. Thus  the material 
gathered by the star(s) during collapse and the 
star formation process had the same contaminating 
metals and in the same proportion -- given by Eq. 
(22) -- as those detected in their photoionized 
volumes.

Low mass galaxies such as \object{I Zw 18} (V\'\i lchez \&
Iglesias-P\'aramo, 1998; Legrand, 2000), \object{II Zw 40},
\object{NGC 4214} (Kobulnicky \& Skillman, 1996), \object{NGC 1569}
(Devost et~al.\@ 1997), \object{Sextans A} and \object{B} (Kniazev
et~al.\@ 2005), in which abundances have been measured in several
locations all present very uniform metal abundances within their
ionized volumes. Given the size of these \hii{} galaxies ($\geq
1$~kpc), these facts imply a very even large-scale dispersal of the
metals from former stellar generations. Scalo \& Elmegreen (2004)
summarize the evidence for abundance variations in the ISM of our
Galaxy, concluding that the gas is well-mixed on scales from 1 to
100~parsec. Note that the inhomogeneities posited in the scenario
  considered in this paper are on much smaller scales.

Abundances in the ISM can also be obtained from 
the absorption lines produced by the \hi\ 
envelopes that surround \hii{} regions.  These 
derivations are difficult and uncertain and 
require very high resolution far-UV spectroscopy. 
The more reliable, recent results in external 
galaxies (e.g. Aloisi et~al., 2003, Lebouteillier 
et~al.\@ 2004) indicate an oxygen abundance in 
the \hi\ regions several times lower than in the 
\hii{} regions (although this is not always the 
case, see Thuan et al.\ 2005). A systematic 
difference between  abundances in  \hii{} and 
\hi\ regions has many possible explanations. One 
of them is that abundance determinations using 
absorption and emission lines have significantly 
different biases in the case of abundance 
inhomogeneities.

Finally, using long-slit spectroscopy of the Orion nebula, Rubin
et~al.\@ (2003) have shown the presence of temperature fluctuations on
a scale of 0.5\arcsec. In the same object, from the ratio of \Oiiit\
and \Oiii\ \textit{Hubble Space Telescope} images, O'Dell et~al.\@
(2003) found small scale temperature fluctuations of $\pm$ 400\,K on
the 10\arcsec\ scale, but with most power being at smallest scales:
0.1\arcsec, i.e., $6.45 \times 10^{14} $cm.  O'Dell et~al.\@ (2003)
say that shadows behind clumps close to the ionization fronts can
partly account for the observed temperature fluctuation, but an
additional cause is needed. As a matter of fact, these observed
fluctuations could be due to our metal-rich droplets. The
  dominant scale of the observed fluctuations is tantalisingly close
  to the droplet sizes derived in Section~2.2. Correlations of these
observed fluctuations with surface brightness would give a hint on
whether our interpretation is correct.

\section{How much does it rain?}
\label{sec:how-much-does}
In order to see what mass fraction of oxygen can 
be expected in the droplets, here we consider a 
simple scheme for the life cycle of metal-rich
droplets assuming a steady state.

\subsection{Formal derivation}
\label{sec:formal-derivation}
If one assumes that all the oxygen ejected by 
massive stars eventually rains down in the form 
of highly metallic droplets, the rainfall rate of 
oxygen in drops at a given time $t$ is given by:
\begin{equation}
R(t)=e_*({\rm O})\, \dot\Sigma_*(t-\Delta t),
\end{equation}
where $e_*({\rm O})$ is the mass fraction of all stars formed that is ejected as 
oxygen by SNe, and $\dot\Sigma_*(t-\Delta t)$ is 
the star formation rate (SFR) per unit area at 
$t-\Delta t$, with $\Delta t\sim10^8$ yr, the 
time needed for SNe ejecta to travel through the 
halo and return to the disk. The droplets end up 
being  homogeneously distributed in the ISM, to 
then be fully processed while enhancing the 
metallicity of newly formed stars and of their 
photoionized surrounding gas, every time that 
they participate in  new bursts of stellar 
formation.

If $X_\mathrm{i}$ is the mass of gas that is 
ionized per unit mass of stars formed, the 
fraction of gas ionized or incorporated into 
stars per unit time is
$I(t) = (X_\mathrm{i}+1)\,\dot\Sigma_*(t)/\Sigma_{\rm 
ISM}(t)$, where $\Sigma_{\rm ISM}$ is
the surface
density of gas. The rate per unit area at which oxygen is released from the
droplets to the ISM is:
\begin{equation}
I(t)\,
        \Sigma_{\rm D}({\rm O};t) = 
\frac{(X_\mathrm{i}+1)\,\dot\Sigma_*(t)}{\Sigma_{\rm 
ISM}(t)}\,\Sigma_{\rm D}({\rm
O};t),
\end{equation}
where $\Sigma_{\rm D}({\rm O};t)$ is the oxygen 
mass in droplets per unit area at
time
$t$. The time variation of $\Sigma_{\rm D}({\rm O};t)$ is then:
\begin{equation}
\frac{\mathrm{d}\Sigma_{\rm D}({\rm 
O};t)}{\mathrm{d}t}=R(t)-I(t)\,\Sigma_{\rm D}({\rm 
O};t).
\end{equation}
If one considers a time interval $t-t_0$ during 
which $\dot\Sigma_*(t-\Delta t)$, 
$\dot\Sigma_*(t)$, and $\Sigma_{\rm ISM}(t)$ are 
approximately constant, eq.\ (9) can be 
integrated to give:
\begin{equation}
\Sigma_{\rm D}({\rm O};t)=\Big(\Sigma_{\rm D}({\rm
O};t_0)-\frac{R}{I}\Big)\,e^{-I(t-t_0)}
        +\frac{R}{I}.
\end{equation}
The drops processing timescale is thus:
\begin{equation}
\tau_{\rm D}=\frac{1}{I}=\frac{\Sigma_{\rm 
ISM}}{(X_\mathrm{i}+1)\dot\Sigma_*(t)}.
\end{equation}
If we take $X_\mathrm{i}=9$ masses of ionized gas 
per unit mass in new stars (Williams \& McKee 
1997), with $\Sigma_{\rm ISM}\simeq10$ M$_\odot$ 
pc$^{-2}$ (see Fig.~16 of Moll\'a \& D\'\i az 
2005) and $\dot\Sigma_*=7.5$ M$_\odot$ pc$^{-2}$ 
Gyr$^{-1}$ (McKee \& Williams 1997) in the solar 
neighbourhood, $\tau_{\rm D}\simeq130$ Myr. And 
for $t-t_0$ equal to 2--3 times the processing 
timescale $\tau_{\rm D}$, we get from eq.\ (10):
\begin{equation}
\Sigma_{\rm D}({\rm O};t)\simeq\frac{R}{I}=
\frac{e_*({\rm O})\,\dot\Sigma_*(t-\Delta 
t)\,\Sigma_{\rm ISM}}{(X_\mathrm{i}+1)\,
        \dot\Sigma_*(t)},
\end{equation}
and if the SFR has been approximately constant 
for the last one or few Gyr (probably a good 
approximation to within a factor of 2 for our own 
Galaxy -- see Fig.~15 of Moll\'a \& D\'\i az 
2005):
\begin{equation}
\Sigma_{\rm D}({\rm O};t)=\frac{e_*({\rm O})\,\Sigma_{\rm ISM}}{(X_{i}+1)}.
\end{equation}

The surface density of gas in the droplets with respect to the surface density of gas in the ISM is 

\begin{equation}
X=\frac{\Sigma_{\rm D}}{\Sigma_{\rm 
ISM}}=\frac{\Sigma_{\rm D}({\rm O};t)}{\Sigma_{\rm ISM}({\rm O};t)} \frac{Z_{\rm ISM}({\rm O})}{Z_{\rm D}({\rm 
O})},
\end{equation}
where $Z_{\rm D}({\rm O})$ and ${Z_{\rm ISM}({\rm 
O}})$ are the oxygen mass fractions in the droplets and in the ISM, respectively.

If we define the parameter $\eta$ by:

\begin{equation}
\eta=
        \frac{e_*({\rm O})}{(X_\mathrm{i}+1)\,Z_{\odot}({\rm O})},
\end{equation}
it follows from Eqs. (13) and (14) that: 
\begin{equation}
X=\frac{\eta}{Z_{\rm D}({\rm O})/Z_{\odot}({\rm O})}.
\end{equation}

In first approximation, the value of $\eta$ is 
not expected to vary strongly with metallicity, 
so the value of $X$ is roughly inversely 
proportional to the metallicity of the droplets.

\subsection{Numerical estimates}
\label{sec:numerical-estimates}
\begin{table}
\caption{Mass fraction of all stars formed that is ejected as oxygen.}
\label{}\centering\begin{tabular}{llcc}
\hline\hline
IMF\rlap{\textsuperscript{a}} & $Z/Z_\odot$ & $e_*({\rm O})$ & Ref \\
\hline
Scalo  & 0.1 & 0.0020--0.0024 & 1 \\
Scalo  & 1   & 0.0022--0.0027 & 1 \\
Scalo  & 0.02 & 0.0040 & 2 \\
Scalo  & 1    & 0.0039 & 2 \\
Salpeter  & 0.1 & 0.0052--0.0062 & 1 \\
Salpeter  & 1   & 0.0056--0.0070 & 1 \\
Salpeter  & 0.02 & 0.0104 & 2 \\
Salpeter  & 1    & 0.0100 & 2 \\
Kroupa  & 0.1 & 0.0092--0.0110 & 1 \\
Kroupa  & 1   & 0.0099--0.0120 & 1 \\
Kroupa  & 0.02 & 0.0184 & 2 \\
Kroupa  & 1    & 0.0177 & 2 \\
\hline
\end{tabular}
\begin{tabular}{l}
1- Woosley \& Weaver (1995); 2- Portinari et al.\ (1998)
\end{tabular}
\begin{tabular}{l}
\textsuperscript{a} The IMF slopes are those of Scalo (1986)
-- as given in \\ Lanfranchi \& Matteucci (2003) --, Salpeter (1955), and \\
Kroupa (2001)
\end{tabular}
\end{table}

Neither the value of $e_*({\rm O})$ nor the value 
of $X_\mathrm{i}$ are well-known. In Table 1, we 
show different estimations of  $e_*({\rm O})$. This is approximately
equal to the mass fraction of oxygen ejected by stars of all masses,
since the contribution of intermediate mass stars is negligible. The values shown in Table 1 are 
based on different sets of calculations of 
stellar yields and on different stellar initial 
mass functions. There is presently no consensus 
on which are the most realistic values and 
chemical evolution models of galaxies have been 
constructed exploring different sets of yields 
and initial mass functions (e.g., Portinari et 
al. 1998, Henry et al. 2000, Chiappini et al. 
2003, Fran{\c c}ois et al. 2004). In Table 1, the 
values of  $e_*({\rm O})$ range between 0.002 and 
0.018. They have been computed for a lower 
stellar mass of  $M_\mathrm{down}=0.1~M_\odot$ 
and an upper stellar mass of 
$M_\mathrm{up}=120~M_\odot$. Adopting larger 
values of $M_\mathrm{down}$ but maintaining the 
same  initial mass functions would lead to 
somewhat larger values of $e_*({\rm O})$ (e.g. 
0.015 for $M_\mathrm{down}=0.3~M_\odot$ instead 
of 0.01 for $M_\mathrm{down}=0.1~M_\odot$ in the 
case of the yields of Portinari et al. 1998 with 
the Salpeter initial mass function).

Concerning  $X_\mathrm{i}$, the situation is 
probably even more complex, because an estimate 
of this parameter involves a representation of 
the structure of the ISM on different scales and 
a knowledge of all the forces at play, in 
addition to the IMF description. Using a simple 
argument of ionization balance, Izotov et al. 
(2006) derive $X_\mathrm{i}$=360/$n$, where $n$ 
is the ionized gas density. So, for a density of 
300 cm$^{-3}$, $X_\mathrm{i}$ would be close to 
1. In a comprehensive study of  the evolution of 
molecular clouds in the Galaxy,  Williams \& 
McKee (1997) give $X_\mathrm{i}=9$ while from the 
work of Franco et al. (1994) one derives 
$X_\mathrm{i} \sim 25$.
The parameter $X_\mathrm{i}$ is likely to change 
with metallicity due to a varying amount of 
ionizing photons absorbed by dust or to a change 
in the hardness of the ionizing radiation field.

Figure 1 shows the variations of $X$ as a 
function of $Z_{\rm D}$ for $\eta = 0.17$, our 
working value in the remaining of the paper 
(which corresponds to $e_*({\rm O}) = 0.01$ and 
$X_\mathrm{i} =9$).

  Finally, note that a change by a given factor 
in the star formation rate between $t-\Delta t$ 
and $t$ would change $X$ by the 
same factor.

\begin{figure}\includegraphics[width=5cm]{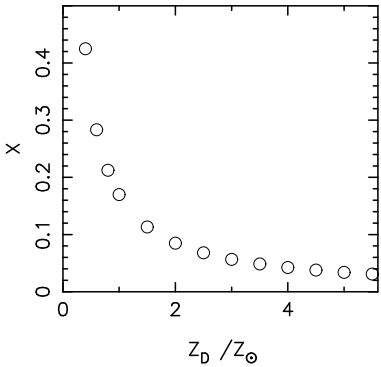}\caption{The 
value of $X=M_{\rm D}/M\ISM$ as a function of the 
metallicity of the droplets, assuming that $\eta$ 
= 0.17 (see Eqs. 18 and 19 in 
text).}\label{fig2}\end{figure}

\section{Metal-rich droplets and the ORL/CEL discrepancy}
\label{sec:metal-rich-droplets-1}
\subsection{ADFs measured in \hii{} regions }
\label{sec:adfs-measured-hii}
\begin{table*}
\caption{Densities, temperatures, metallicities and ADFs measured in \hii{} regions}
\label{}\centering\begin{tabular}{rrrrrrrr}
\hline\hline
ID & $n_\mathrm{e}$ 
(cm$^{-3}$)\rlap{\textsuperscript{a}} & 
$T_\mathrm{e}[\ion{O}{iii}]$ (K) & $T_\mathrm{e}$(BJ) (K) &
        $12+\log(\mathrm{O/H})$\rlap{\textsuperscript{b}}& 
$Z/\Zs$\rlap{\textsuperscript{c}}& 
$\mathrm{ADF}(\mathrm{O}^{++})$ & Ref\\
\hline\multicolumn{7}{c}{Galactic \ion{H}{ii} regions} \rule[-4pt]{0pt}{12pt}\\
\hline
\object{M8}  & 1750 & $8050\pm700$ & -- & 8.41 & 0.52 & 2.0 & 1 \\
             & 1800 & $8090\pm140$ & 7100$^{+1250}_{-1000}$ & 8.51 & 0.66 & 2.3 & 2 \\
\object{M16} & 1120 & $7650\pm250$ & $5450\pm820$ & 8.56 & 0.74 & 2.8 & 3 \\
\object{M17} & 860, 520 & $8120\pm250$, $8210\pm250$ & -- & 8.53, 8.50 & 0.69, 0.65 &1.8, 2.2 & 4 \\
              & 600--1500 & 8200 & 7700 & 8.56 & 0.74 & 2.1 & 5 \\
              & 470 & 8$020\pm170$ & -- & 8.52 & 0.68 & 1.9 & 2 \\
\object{M20} & 270 & $7800\pm300$ & $6000\pm1300$ & 8.53 & 0.69 & 2.1 & 3 \\
\object{M42} & 4000, 5700 & $8300\pm210$, $8350\pm200$ & $8730\pm800$, $8390\pm800$ & 8.47, 
8.47 & 0.60, 0.60 & 1.3, 1.5 & 6 \\
              & -- & -- & -- & 8.52 &  0.68 & 1.3 & 5 \\
              & 8900 & $8300\pm40$ & $7900\pm600$ & 8.51 & 0.66 & 1.4 & 7 \\
\object{NGC 3576} & 1300--2700 & 8850 & 8070 & 8.52 & 0.68 & 1.8 & 5 \\
                   & 2800 & $8500\pm50$ & $6650\pm750$ & 8.56 & 0.74 & 1.9 & 8 \\
\object{NGC 3603} & 5150 & $9060\pm200$ & -- & 8.46 &   0.59 & 1.9 & 3 \\
\object{S311} & 310 & $9000\pm200$ & $9500\pm900$ & 8.39 & 0.50 & 1.9 & 9 \\
\hline\multicolumn{6}{c}{Extragalactic \ion{H}{ii} regions}\\
\hline
LMC \object{30 Dor} & 370--1800 & 10100 & -- & 8.34 & 
0.45 & 2.0--2.7\rlap{\textsuperscript{d}} & 5 \\
                     & 316 & $9950\pm60$ & $9220\pm350$ & 8.33 & 0.44 & 1.6 & 10 \\
\object{LMC N11B}   & 80--1700 & 9400 & -- & 8.41 & 
0.52 &4.9--8.2\rlap{\textsuperscript{d}} & 5 \\
\object{NGC 604}    & $\le100$ & $8150\pm150$ & -- & 8.49  & 0.63 & 1.6 & 11 \\
\object{NGC 2363}   & 360--1200 & $15700\pm300$ & -- & 7.87 & 0.15 & 2.2 & 11 \\
\object{NGC 5253}   & 370--610 & 10940--12010 & -- &
8.18--8.28 & 0.31--0.39 & 1.5--1.9 & 12 \\
\object{NGC 5461} & 300 & $8600\pm250$ & -- & 8.56 & 0.74 & 1.9 & 11 \\
\object{NGC 5471} & 220--1150 & $14100\pm300$ & -- & 8.03 & 0.22 & 1.6 & 11 \\
\object{NGC 6822 V} & 175 & $11900\pm250$ & -- & 8.08 & 0.24 & 1.9 & 13\\
\object{SMC N66} & 50--3700 & 12400 & -- & 8.11 & 0.26& 2.3 & 5 \\
\hline
\end{tabular}
\begin{tabular}{l}
1- Esteban et al.\ (1999a); 2- Garc\'\i a-Rojas et al.\ (2006b);
3- Garc\'\i a-Rojas et al.\ (2006a); 4- Esteban et al.\ (1999b);
\\
5- Tsamis et al.\ (2003); 6- Esteban et al.\ (1998);
7- Esteban et al.\ (2004); 8- Garc\'\i a-Rojas et al.\ (2004);\\
9- Garc\'\i a-Rojas et al.\ (2005); 10- Peimbert (2003);
11- Esteban et al.\ (2002); 12- L\'opez-S\'anchez et al.\ (2007);\\
13- Peimbert et al.\ (2005)
\end{tabular}

\begin{tabular}{l}
\textsuperscript{a} The high values of the 
densities in the extragalactic  \hii{} regions 
mostly come from the \ariv\ ratio and are very 
uncertain\\\textsuperscript{b} 
$12+\log(\mbox{O/H})$ from forbidden lines 
\\\textsuperscript{c} the Solar oxygen abundance 
is taken from Allende Prieto et~al.\  (2001) 
\\\textsuperscript{d} In 30 Dor and N11B, the 
highest values of the ADFs include a correction 
for
contamination by scattered light
\end{tabular}
\end{table*}

Table 2 lists the values of  the densities, temperatures and  abundance 
discrepancy factors ADF(\Opp) found in the 
literature for Galactic and extragalactic \hii{} 
regions. The temperatures derived from Balmer jump, $T_\mathrm{e}$(BJ), are not available for all cases and bear generally significant uncertainties. It is however clear that they are, on average, smaller than those derived from \rOiii\ line ratios, $T_\mathrm{e}[\ion{O}{iii}]$. The listed ADFs are equal to the ratio of the value of the \Opp\ abundance derived from \Oiii\ to the one derived from  \ion{O}{ii}  recombination lines, using  $T_\mathrm{e}[\ion{O}{iii}]$ in both cases.
These ADFs were obtained from several 
recombination lines of O~{\sc ii} and the 
forbidden line \Oiii,
using the temperature  \TrOiii\ derived from the 
\rOiii\  ratio. The values of ADF(\Opp) are 
found to be  around 2,
over a large range of \hii{} region 
metallicities: $12 + \log (\mathrm{O/H}) = 
7.87$--8.56 (i.e., $Z= 0.15$ to $0.75 \Zs$). 
However, note that the ADFs measured for most of 
the extragalactic \hii{} regions in Table 1 and, 
in particular, for all the low metallicity 
objects (12 + log (O/H) $< 8.3$) were derived 
using the \ion{O}{ii} lines of multiplet 1, and 
these lines can be severely affected by stellar 
absorption lines (Tsamis et al.\ 2003). This 
means that the ADFs could be much larger than 2 
for these objects.

\subsection{A simple expression for the ADF}
\label{sec:simple-expr-adf}To better understand what an ADF really is, we 
derive here an approximate expression for it in 
the case of a two-component toy model.
In a medium characterized by an  electron density 
$n_{\rm e}$ and an  electron temperature $T_{\rm 
e}$, the luminosity
in an optically thin  line  arising from \Opp\  ions is given by
\begin{equation}
L= N({\rm O}^{++}) n_{\rm e} \epsilon(T_{\rm e}),
\end{equation}
where $N$(\Opp) is the total number of  \Opp\ 
ions in the emitting volume, and $\epsilon(T_{\rm 
e})$ is the line
emissivity, in erg cm$^3$ s$^{-1}$, which depends 
essentially on the electron temperature (for the 
sake of simplicity, we neglect the density 
dependence, as it is not important in the problem 
considered here). The expression for 
ADF(O$^{++}$) is then given by
\begin{equation}
{\rm ADF(O}^{++}) = \frac{L_{\rm ORL}/  \epsilon 
_{\rm ORL}(T_{\rm e})}  {L_{{\rm CEL}}/ \epsilon 
_{{\rm CEL}}(T_{\rm e}) },
\label{}
\end{equation}
where  $T_{\rm e} $ is taken equal to \TrOiii. However,
if within the emitting volume there are two 
different  media with different physical 
parameters such as their metallicity, then
\begin{equation}
{\rm ADF(O}^{++}) = \frac{(L^a_{\rm ORL}+ 
L^b_{{\rm ORL}})/\epsilon_{\rm ORL}(T_{\rm e})}
{(L^a_{\rm CEL}+ L^b_{\rm CEL})/\epsilon_{{\rm CEL}}(T_{\rm e})},
\label{}
\end{equation}
where  $a$ and $b$ refer to the two  media, and
$T_{\rm e}$ is here again equal to the 
temperature derived from the observations, 
\TrOiii.
In the limiting case in which the  medium $b$ is 
too cold  to  contribute significantly  to the 
CEL intensities (which is
easy to achieve due to the very strong 
temperature dependence of the $\lambda 4363$  and 
$\lambda 5007$
lines ), \TrOiii\ is in fact the temperature of
medium $a$. Using Eq. (17), Eq. (19) becomes:
\begin{equation}
{\rm ADF(O}^{++}) = 1+\frac{N^b({\rm 
O}^{++})}{N^a({\rm O}^{++})} \frac{n^b_{\rm e} }
{n^a_{\rm e} } \frac { \epsilon_{\rm ORL}(T^b_{\rm e})}
{\epsilon_{\rm ORL}(T^a_{\rm e})} ,
\end{equation}
or,
\begin{equation}
{\rm ADF(O}^{++}) = 1+\frac{x^b({\rm O}^{++})}{x^a({\rm O}^{++})}
\frac{n^b_{\rm e} }{n^a_{\rm e} }
\frac{Z^b}{Z^a}
\frac{M^b}{M^a}
  \frac { \epsilon_{\rm ORL}(T^b_{\rm e})}
{\epsilon_{\rm ORL}(T^a_{\rm e})},
\label{}
\end{equation}
where $x^a$, $x^b$ stand for the doubly ionized 
oxygen fractions, $Z^a$, $Z^b$ are the O/H ratios 
(or ``metallicites''),
and $M^a$, $M^b$ are the total hydrogen masses in zones
$a$ and $b$, respectively. Note that 
$\epsilon_{\rm ORL}$ is an increasing function of 
$Z$ via $T_{\rm e}$, and $x({\rm O}^{++})$ is a 
decreasing function of  $n_{\rm e}$ due to 
recombination. The observed values of the ADFs in 
\hii{} regions impose strong
constraints on the parameters that come into play.

\subsection{Photoionization modelling}
\label{sec:phot-modell}
\begin{figure}\includegraphics[width=5cm]{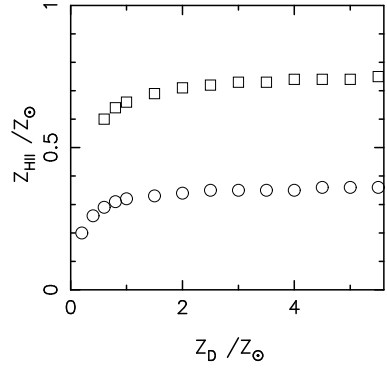}\caption{The 
metallicity of the fully mixed \hii{} region as a 
function of the metallicity of the droplets, for 
$Z_{\rm ISM} / Z_{\odot}$ = 0.2 (circles) and 
$Z_{\rm ISM} / Z_{\odot}$ = 0.6 (squares), 
assuming that $\eta$ = 0.17 (see Sect. 3.1).}\label{fig2}\end{figure}

We now simulate the situation described in Sect. 
2 by a set of photoionization models. We consider 
three ionized components, each with its own 
metallicity:
the highly metallic droplets, the background ISM, which is being ionized at
the same time as  the droplets, and the
\hii{} region into which the former two media have mixed. At each moment, a
mass $M_{\rm D}$ of droplets with metallicity 
$Z_{\rm D}$ and a mass $M\ISM$ from the ISM with 
metallicity $ Z\ISM$
are being ionized and will eventually result in 
a mass  $M_{\rm HII} $ of ionized gas with 
metallicity
\begin{equation}
Z_{\rm HII}= \frac{X Z_{\rm D} +Z\ISM}{X+1}  ,
\end{equation}
where $X=M_{\rm D}/M\ISM$. Figure 2 shows $Z_{\rm 
HII}$ as a function of $Z_{\rm D}$ for $Z\ISM$ = 
0.2 and 0.6 $Z_{\odot}$, taking $\eta$ = 0.17.

The real progression of the ionization front is 
complex, but the situation can be modelled in a 
simple way by
considering that the ISM, the droplets and the fully mixed \hii{} region,
with typical densities $n\ISM$,  $n_{\rm D}$ and 
$n_{\rm HII}$, respectively,  are impacted by the 
same ionizing radiation field.
This implies that the droplets do not have too large
a covering factor so that the general ISM can be 
ionized (in principle, a situation where the 
droplets would completely block
the ionizing photons is possible as well, but not considered in this paper).

Here we assume that any observation of an \hii{} 
region encompasses zones close to the ionization 
front that have
just been ionized and still have their initial 
metallicities $Z_{\rm D}$ and $Z\ISM$, and zones 
that have been
ionized some time ago and have acquired, through 
the turbulent stirring promoted by well localized 
champagne flows (see Sect. 2),
the metallicity $Z_{\rm HII}$ given by Eq. (22).
We assume that this fully mixed
gas contributes a fraction, $F$, of the total observed H$\beta$ flux
from the \hii{} region. Note that the parameter 
$F$ also accounts for the lifetime of the 
droplets once they are ionized.

For a given stellar radiation field, and a given 
value of $\eta$ -- as defined by Eq. (15) -- the problem is entirely 
specified by the parameters $F$, $ Z_{\rm D}$ 
and $ Z\ISM$,  and the densities of the three 
different media. We expect the densities $ n\ISM$ 
and $ n_{\rm HII}$ to be of the same order of 
magnitude because  the fully mixed ionized gas must be in rough pressure equilibrium
with  the ISM gas that has just been ionized and because 
temperature differences between the media are likely small, 
which is confirmed by our results shown in Figs. 4$d$ and 5$d$.  
For the computations, inspired by the 
observed values listed in Table 2, we have taken 
$ n\ISM =  n_{\rm HII} = 300 \mathrm{\ cm}^{-3}$. 
Since  the droplets are more metallic than the 
ISM, their  temperature must  be lower, and thus 
we assumed the droplets to be denser than the 
background. We have considered various density 
contrasts and present results from calculations 
assuming $ n_\mathrm{D} = 1000 \mathrm{\ 
cm}^{-3}$. The computations were done  with the 
code PHOTO (Stasi\'nska 2005), taking for the 
ionizing radiation field that of a blackbody at a 
temperature   $T_{\rm eff}$ = 50,000~K. The inner radius of the 
models is the same for the three components, and 
has been chosen so that  the  ionization 
parameter at the inner radius, $U$,  is equal to $10^{-2}$ both for 
the \hii{} region and the ISM\footnote {We use 
the definition $U = Q_{\rm H }/ (4 \pi R^2 n)$, 
where $Q_{\rm H}$ is the number of stellar 
ionizing photons, $R$ is the distance to the 
star, and $n$ is the gas density.}.  Fig. 3 shows 
how  the \Opp-weighted temperature in the 
droplets decreases as a function of the droplet 
metallicity.
\begin{figure}\includegraphics[width=5cm]{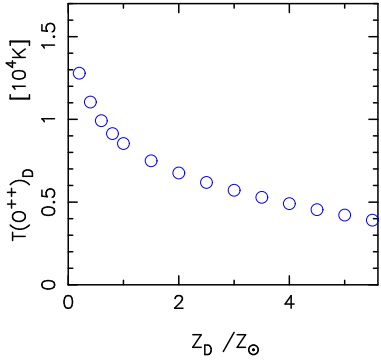}\caption{The 
average temperature in the \Opp\ zone of the 
droplets, as a function of the metallicity of the 
droplets.}\label{fig2}\end{figure}

We computed models of the droplets with various values of $Z_{\rm
    D}$, and models for the mixed \hii{} region with $Z_{\rm HII}$ as
  given by Eq. (22), the value of $X$ resulting from Eq. (16) for an
  adopted $\eta$ of 0.17. We combined the photoionization models for
  the droplets, the ISM and the fully mixed \hii{} region taking
  various values of $F$. We then applied to the resultant models the
same techniques as an observer deriving \TrOiii, ADF(\Opp), and the
metallicities from the collisionally excited lines and recombination
lines.  We used exactly the same atomic data as for the
photoionization models.

\subsection{Discussion of the simulated observations}
\label{sec:disc-simul-observ}
The number of parameters in this problem is large, and displaying our
results for the entire parameter space would be cumbersome.  Since the
main goal of this paper is to investigate whether the ADFs observed
in \hii{} regions could indeed be the signature of the enrichment
scenario proposed by T-T96, we will concentrate on a few plausible
cases, and try to learn some lessons from them.

\begin{figure*}
   \includegraphics[width=17cm]{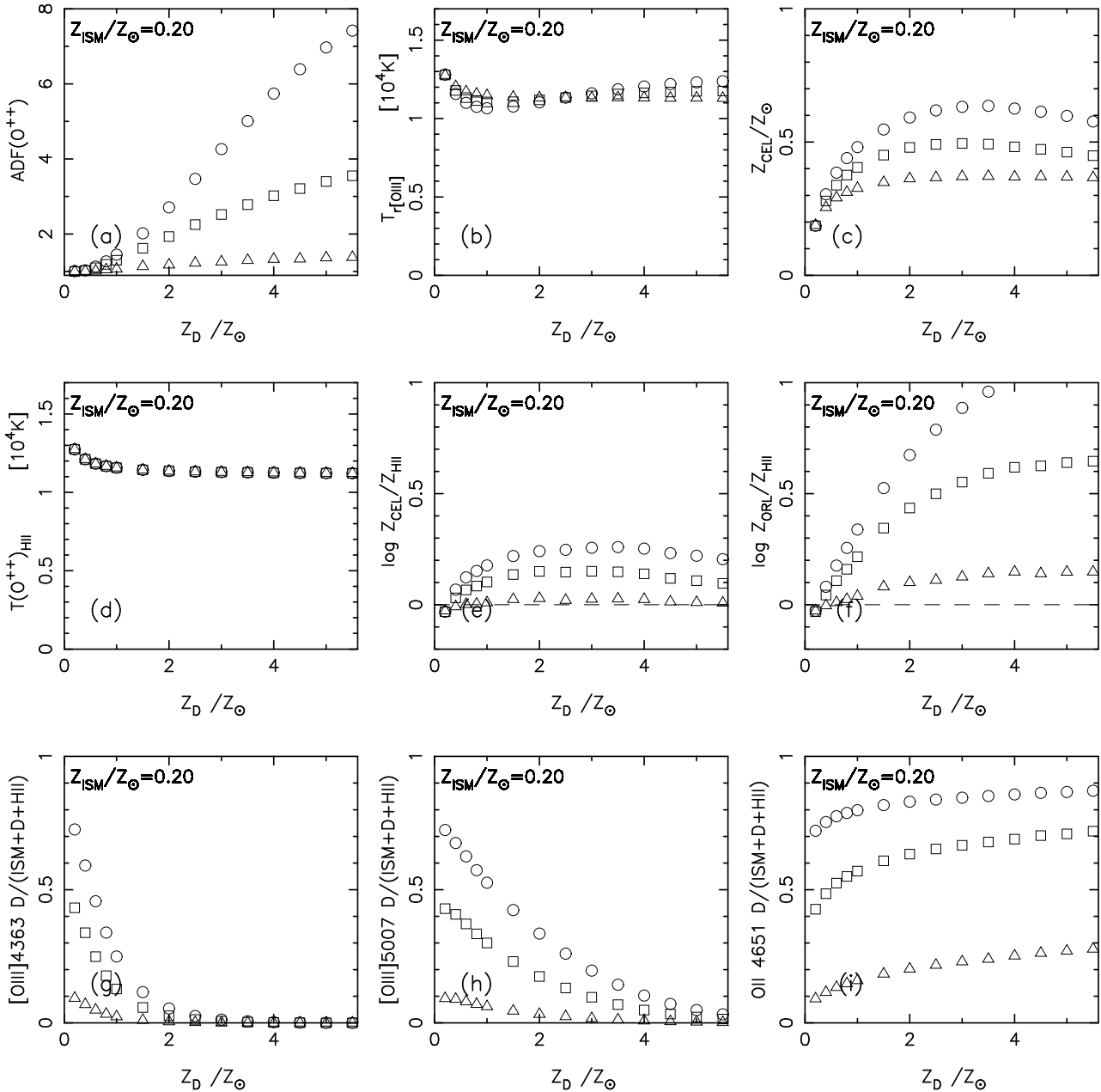}
   \caption{Results of simulations for a series of models with
     $Z\ISM=0.2~\Zs$ as a function of $ Z_{\rm D}$, for various values
     of $F$ and taking $\eta$=0.17 (see Sect.3.2 and
       4.3). Circles: $F$=0.1; squares: $F=0.5$; triangles: $F=0.9$.
     ``Observables'': (\textit{a})~ADF(\Opp), (\textit{b})~\TrOiii,
     (\textit{c})~$Z_{\rm CEL}$ (see text). Other results of the
     simulations: (\textit{d})~$T$(\Opp)$_{\rm HII}$,
     (\textit{e})~$Z_{\rm CEL}$/$Z_{\rm HII}$, i.e., the metallicity
     bias when using \Oiii\ (see text) (\textit{f})~$Z_{\rm
       ORL}$/$Z_{\rm HII}$, i.e., the metallicity bias when using the
     \ion{O}{ii}~$\lambda$ 4651 line to measure the oxygen abundance
     (see text); (\textit{g}), (\textit{h}) and (\textit{i}): the flux
     from the droplets in the \Oiiit, \Oiii\ and \ion{O}{ii}
     $\lambda$4651 lines, with respect to the total in these lines.}
   \label{fig1}
\end{figure*}

\begin{figure*}
   \includegraphics[width=17cm]{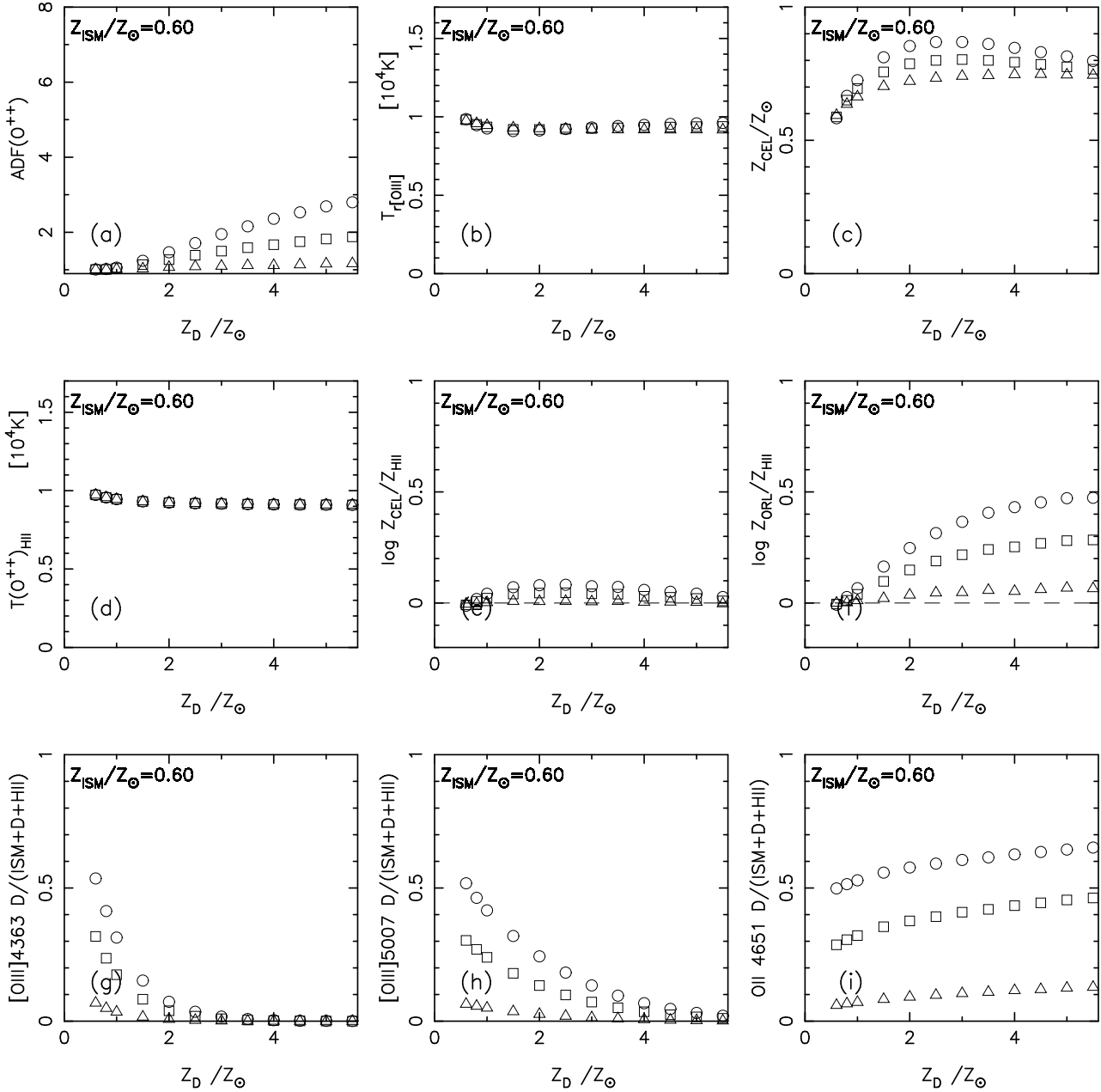}
   \caption{Same as Fig. 2 but for models with $Z\ISM=0.6~\Zs$.}
   \label{fig2}
\end{figure*}

Figure 4 presents the results of  simulations 
for $ Z\ISM=0.2~\Zs$, taking $\eta$=0.17 (the 
working value presented in Sect. 3.2). In all the 
nine panels   the abscissa represents $ Z_{\rm 
D}$   in solar units. The range of values for $ 
Z_{\rm D}$ was inspired by the work  of Silich 
et~al.\@ (2001) where the metallicity of the hot 
superbubble interior, resultant from the multiple 
SN explosions of an aging star cluster, was 
calculated. The various cases they considered 
(see their Fig. 6) encompass starbursts evolving 
in low and high metallicity galaxies and account 
for mass evaporation  from the surrounding 
supershell into the hot superbubble interior. The 
final values, the metallicity of the matter that 
will eventually cool to compose the high 
metallicity droplets, range from 0.2~\Zs{} to 
5~\Zs. In the plots, simulations  corresponding 
to an observation where $F=0.1$ are represented 
by circles, those corresponding to $F=0.5$  by 
squares, and those  corresponding to $F=0.9$ by 
triangles. Panel \textit{a} shows the simulated 
ADF(\Opp), panel \textit{b} shows  \TrOiii, 
panel \textit{c} shows $Z_{\rm CEL}$, the 
metallicity derived from collisionally excited 
lines.  These three quantities represent 
``observables''. The next panels show other 
results from the simulations, which shed some 
light on  the behaviour of the ``observables''. 
Panel \textit{d} shows $T(\Opp)_{\rm HII}$, the 
\Opp-weighted  temperature in the fully mixed 
\hii{} region (which, as can be seen by comparing panels b and d, is not identical to the temperature derived from the {\rOiii} ratio). Panels \textit{e} and \textit{f} 
show the bias in the determination of the 
metallicity when using collisionally excited 
lines  ($Z_{\rm CEL}$/$Z_{\rm HII}$) or 
recombination lines ($Z_{\rm ORL}$/$Z_{\rm 
HII}$). The computed values of $Z_{\rm CEL}$ and 
$Z_{\rm ORL}$ take into account the contribution 
of \Op, which has been evaluated from the \Oii\ 
line using the temperature derived from the 
\rNii\ line ratio. Note that this procedure gives 
lower ORL abundances than the procedure used by 
most authors quoted in Table 2, who correct the 
temperature of the \Op\ zone for temperature 
fluctuations using the scheme of Peimbert (1967). 
Panels \textit{g}, \textit{h} and \textit{i} show 
the flux from the droplets in the  \Oiiit, 
\Oiii{} and  \ion{O}{ii} $\lambda$4651 \footnote {Following the usual nomenclature for recombination lines, the \ion{O}{ii} $\lambda$4651 line is emitted by the recombining \Opp\ ion.} lines, 
with respect to the total in these lines.

As expected, ADF(\Opp) increases with $ Z_{\rm 
D}$, and, for a given $ Z_{\rm D}$, it is larger 
for smaller values  of $F$. The computed $Z_{\rm 
CEL}$ also strongly depend on $F$.  In other 
words, the computed metallicity is higher if the 
observation gathers less light from the fully 
mixed  \hii{} region with respect to the  newly 
ionized gas. The behaviour of  \TrOiii\ is more 
intricate.  As $ Z_{\rm D}$ increases, the 
temperature in the drops becomes smaller, as seen 
in Fig. 3, due to the increased cooling by 
metals. One would expect \TrOiii\  to decrease as 
$ Z_{\rm D}$ increases. This is seen only at 
small values of $ Z_{\rm D}$. The behaviour of 
\TrOiii\  is actually due to a subtle combination 
of various effects, linked to the temperatures in 
the three phases of our models and in the 
respective proportions of these phases in the 
resultant model. The temperature in the fully 
mixed \hii{} region decreases slightly as $ 
Z_{\rm D}$ increases, because $ Z_{\rm HII}$ 
increases  as prescribed by Eq.~(22), which 
accounts for mass and metal conservation. 
However Fig.~4\textit{g} shows that the drops 
contribute less to the \Oiiit\ flux than to the 
\Oiii\ flux, and thus, the \rOiii\ ratio is 
larger when the contribution of the resultant 
\hii{} region to \Oiii\ is smaller. The value  of 
$Z_{\rm CEL}$ is not only related to the ADF, but 
also to the contribution of the \Op\ zone. 
Therefore, the behaviour of $Z_{\rm CEL}$ with  $ 
Z_{\rm D}$ seen in Fig. 4 \textit{c} is not 
necessarily universal. A wide range of numerical 
models, with a better representation of the 
physical model and a better sampling of 
excitations conditions would be needed to draw 
some useful conclusions from the computed values 
of $Z_{\rm CEL}$.

Figure 5 is the same as Fig. 4, but now for 
simulations with $ Z\ISM$=0.6~\Zs. Qualitatively, 
it displays the same characteristics as Fig. 4, 
but with less extreme features. The  ADFs(\Opp) 
are here much smaller, because the temperature 
contrast between the metal-rich drops and the ISM 
is smaller.
On the other hand, simulations with $ Z\ISM$ 
smaller than 0.2~\Zs, not shown here, lead to 
more extreme characteristics than displayed in 
Fig. 4.

The same set of figures  for droplets with a 
density $ n_{\rm D}$ = 3000 instead of 1000 
cm$^{-3}$ are very similar to Figs. 4 and 5,  but 
with smaller effects, due to the fact that gas is 
more recombined in the droplets than in the cases 
shown here. However, the density effect is only 
slight, since, as seen in Eq.~(21) the effects of 
density and recombination on ADF(\Opp) tend to 
compensate.

 Our computations were carried out with a given stellar radiation field
and with a given ionization parameter. We do not expect large changes
in our results when modifying the stellar radiation field or
ionization parameter within the limits relevant to bright \hii{}
regions. The average effective temperature does not change much from
one object to another, and is a second order parameter in our
problem. The ionization parameter can be different from object to
object, but the ADFs are abundance ratios of the same ion, and thus
are not dependent on the ionization parameter to first order. Test
computations have confirmed this view.

The most important parameters are $ Z_{\rm D}$ and $F$ (which are
varied in the plots presented in Figs. 4 and 5) and our parameter
$\eta$, which, from considerations in Sect. 3.2, has a large range of
possible values.

Table 3 summarizes our knowledge on the range of parameters entering the model.

\begin{table}
\caption{Expected range of parameters entering the model.}
\label{}\centering\begin{tabular}{lll}
\hline\hline
parameter                & values           				&  argumentation  \\
\hline
$e_*({\rm O})$       & 0.002 -- 0.018                    &  Table 1  \\
$X_{\rm i} $         & 1 ? -- 25 ?                     &  see  Sect. 3.2  \\
$ Z_{\rm D}$         & $ Z\ISM$ -- 10\Zs ?             & Silich et al. (2001) \\
$n_{\rm HII}$       &  100 -- 5\,000$\mathrm{\ cm}^{-3}$  & Table 2 \\
$ n\ISM $  /$ n_{\rm HII}$          & $ \simeq $ 1    &  see Sect. 4.3            \\ 
$ n_\mathrm{D}$ /$ n\ISM $    &  3  -- 30             &   rough pressure equilibrium           \\
$U$                  & $ 10^{-3}$ -- $10^{-1} $  &   Stasi\'nska \& Leitherer (1996)              \\
$T_{\rm eff}$        & 35,000 -- 55,000~K      &     massive stars \\
$F$                  & 0 -- 1                    &  see Sect. 4.4  \\
\hline
\end{tabular}
\end{table} 

\subsection{What the confrontation between computed ADFs and observed ADF tells us}
\label{sec:what-confr-betw}
Let us now compare the simulations shown in Figs. 
4 and 5  with the observations.  The Galactic and 
Magellanic Clouds nebulae listed in Table 2 have 
diameters of $(10-50)\arcmin$ while the slits 
have widths of  $(1-3) \arcsec$,  so that the 
spectroscopic observations cover a tiny fraction 
of the objects, and the contribution from the 
fully mixed   \hii{}  region is likely small. The 
remaining extragalactic objects are smaller, with 
angular radii of $(10-100)\arcsec$, so the 
contribution of the fully mixed \hii{}  region is 
more important, however still small. We thus 
believe that the models to be compared to the 
available observations are those corresponding to 
the smallest values of $F$.  The ADFs  listed in 
Table 2 are mostly in the range 1.5--2, with only 
few cases above 2. Such values are obtained for 
our models for a variety of physical conditions.

Our models show clear trends in the values of 
ADFs with $ Z\ISM$, while there is no obvious 
trend of the observed ADFs with any observed 
physical parameter.
Note, however,  that, for the extragalactic 
\hii{} regions, the uncertainties in the 
absorption corrections (and whether they should 
be applied) and the fact that for most of the 
extragalactic objects the ADF has been determined 
from just a couple of lines (or line-blends) of 
multiplet 1 (which are very weak and difficult to 
measure), preclude us from excluding any of the 
following possibilities: \emph {i)} all the 
extragalactic  \hii{}  regions show ADFs~$\sim2$; 
\emph {ii)} some extragalactic  \hii{} regions 
have high ADFs; \emph {iii)} the low-metallicity 
\hii{}  regions show higher (lower) ADFs. From 
our models, low metallicity objects are bound to 
show higher ADFs. Obviously, a more conclusive 
sample of observational data is needed to be 
confronted with our model predictions.

It has been commented by L\'opez-S\'anchez et al. 
(2007) that observed ADFs are similar in a 
variety of  \hii{}  regions in different host 
galaxies. They take this fact as an argument 
against metal-rich droplets producing the ADFs. 
The entire grid of our models does predict  a 
wide range of ADF values, which seems in 
contradiction with the very restricted range of 
observed ADFs and the fact that there seems to be no 
correlation between them and any observed property of the  \hii{} 
regions (see Garc\'ia-Rojas \& Esteban 2006). However, not all our models 
necessarily represent a physically plausible 
model. For example, the value of  $ Z_{\rm D}$ is 
certainly constrained by physical processes 
occuring in the superbubbles, so that only a 
fraction of the considered range is relevant. In 
this respect, we might even suggest that, in the 
framework of our scenario to explain the ADFs, 
the observed values could put constraints on the 
mixing of metals in the supernova bubbles.
 For example, the observed ADFs do not seem to depend on metallicity
whereas our models would predict larger ADFs for a lower $Z\ISM$ if
all other parameters were to remain fixed. From Table 2, the values of
$e_*({\rm O})$ do not change with metallicity. Thus, if our scenario
is to explain the observed ADFs, this imples that one or more of the
remaining parameters ($F$, $X_\mathrm{i}$, or $ Z_{\rm D}$) must
correlate with $Z\ISM$ in such a way as to maintain a constant ADF.
Indeed, one might expect a relation between $X_\mathrm{i}$ and
$Z\ISM$, via metallicity-dependent stellar winds which play an
important role in the shaping of the ISM and therefore on determining
the mass of gas that is ionized. One might also expect a relation
between $ Z_{\rm D}$ and $Z\ISM$, although Silich et al. (2001) show 
that this relation is not a simple one.

Note that the magnitude of the biases found in 
our simulations obviously depend on the value 
adopted for $\eta$. Had we used a value of 
$\eta$=0.013, a minimum value implied by the 
references listed in Sect. 3.2, the biases would 
have been very small in most cases.

\subsection{Possible effects of dust}
\label{sec:poss-effects-dust}
In our models we did not consider the effect of 
dust. This was on purpose, in order not to 
inflate the number of free parameters. One of the 
expected effects of dust would be to decrease the 
amount of O available to produce recombination 
lines in droplets  if a significant fraction of 
the O atoms in drops are depleted onto refractory 
dust grains. However, maximum dust condensation 
efficiencies in SNe ejecta seem to be $\sim0.2$ 
(Todini \& Ferrara 2001; Edmunds 2001), so the 
effective value of $X(\rm O)$, would only 
decrease to by a factor of $0.8$. We believe that 
the main effects of dust grains in this problem 
is their contribution to  heat the gas. The 
efficiency of grain heating depends essentially 
on the dust-to-gas mass ratio, the ionization 
parameter, and the size of the grains. The 
contribution of grains to the total heating 
budget is larger for higher ionization parameters 
(Stasi\'nska \& Szczerba 2001). Since the 
droplets are to be denser than the surrounding 
gas, the ionization parameter there is lower, and 
therefore the effect of grain heating is expected 
to be smaller than in the surrounding gas. Thus, 
the effect of the ionization parameter alone 
would be to enhance the contrasts between 
droplets and the rest, and thus to enhance the 
effects discussed in the preceding section. 
However, the properties of the dust grains are 
likely very different in the droplets and in the 
interstellar medium, since the dust in the 
droplets should be heavily weighted by the 
conditions during the supernova ejection and 
droplet condensation. This issue obviously merits 
further study.

\subsection{The abundance bias}
\label{sec:abundance-bias}
It is often said that abundances derived from 
recombination lines are not biased, because all 
recombination lines have roughly the
same temperature-dependence (e.g., Esteban 
et~al.\@ 2005). On the other hand, abundances 
derived from collisionally excited lines could be 
underestimated if large temperature fluctuations 
in the \Opp\ zone are present but not accounted 
for. Some authors apply various techniques to try 
to estimate temperature fluctuations in a given 
object and correct for them using the scheme of 
Peimbert (1967).  Others do not apply such a 
correction, considering that temperature 
fluctuations are expected to be small.

In the case of our scenario, the \emph{real} 
metallicity is that of the fully mixed \hii{} 
region.
Figures  4 \textit{e} and 5 \textit{e} show that 
the bias  $Z_{\rm CEL}$/$Z_{\rm HII}$ is larger 
than 1 in all the cases shown here. Obviously, in 
our scenario, one cannot expect the abundance 
measured from collisionally excited lines to 
measure the  \emph{real}  metallicity exactly. 
The discussion of the abundance bias is 
unfortunately complicated by the fact that one 
should also consider the \Op\ zone, and that the 
results are expected to strongly depend on the 
excitation of the object. We note that, in our 
models, the bias $Z_{\rm CEL}$/$Z_{\rm HII}$ is 
usually small, but it may reach 0.2 -- 0.3 dex if 
the weight of the well-mixed \hii{} region is 
small and if   $ Z_{\rm D}$ is of the order of 2 
$Z_{\odot}$. It is interesting to note that this 
bias is  opposite to the bias due to the 
temperature fluctuations postulated by Peimbert 
(1967). However, many more numerical experiments 
would be needed to consider this as a general 
rule.

Figures 4 \textit{f} and 5 \textit{f} show that 
the bias   $Z_{\rm ORL}$/$Z_{\rm HII}$  is always 
larger than one, and may reach quite large 
factors. As expected, the bias increases with 
ADF(\Opp).

To summarize, if the droplet scenario is valid, ORLs
overestimate the \emph{real} metallicity of the ISM, possibly by quite
large amounts. CELs may also overestimate it, although to a much
weaker extent. The biases are stronger for larger ADFs. Unfortunately,
we have at present no recipe to correct for the biases. In the
meantime, a prudent attitude would be to discard objects with large
ADFs as probes for chemical evolution studies of galaxies.

\section{Conclusions and future directions}
\label{sec:concl-future-direct}

In this paper, we have described a scenario for the enrichment
of the ISM and discussed its observational signatures. The scenario
invokes metal-rich droplets produced by a long process following
supernova explosions in previous stellar generations.  The full
enrichment of the ISM is only achieved when these droplets are
  destroyed by diffusion in photoionized regions. Following Larson
(1996), if the total star formation rate in the galaxy is about
3~$M_\odot$ yr$^{-1}$ and the amount of cloud matter that can be
ionized per every solar mass converted into stars amounts to $\sim
10~M_\odot$ , then if the total gaseous mass of the galaxy amounts to
$4\times 10^9~M_\odot$, the ionization rate implies that the entire
ISM is cycled through the ionized phase every 100 Myr.  It is then
within such a short timescale that the ISM acquires, at least locally,
an even metallicity.  On the other hand, the time required to disperse
the metals from a stellar generation is much longer (several $\times
10^8$ yr, due mostly to a slow cooling process) and a significant
fraction of the mass converted into stars in the last few stellar
generations, which ends up being violently reinserted into the ISM,
remains hidden within the hot phase of the ISM.

We have shown in Section~\ref{sec:metal-rich-droplets-1} that
photoionization of highly metallic droplets can, under certain
conditions, reproduce the observed abundance discrepancy factors
(ADFs) derived for Galactic and extragalactic \hii{} regions.  In
  this scenario the recombination lines arise from the highly
  metallic droplets that have not yet fully mixed with the ISM
  gas. However, we find in Section~\ref{sec:surv-dropl-hii}
that the droplets in \hii{} regions can only have a narrow range of
sizes, bounded from below by the need to avoid rapid destruction and
from above by the need to have eluded detection by direct
imaging.

We find in Section~\ref{sec:abundance-bias} that, if this
scenario holds, the recombination lines strongly overestimate the
metallicities of the fully mixed \hii{} regions.  The collisionally
excited lines may also overestimate them, although in much smaller
proportion (less than 0.1 dex in most of the cases considered in this
study).  The effect increases with ADF(\Opp). In absence of any recipe
to correct for these biases, we recommend that objects showing
large ADFs should be discarded when probing the chemical
evolution of galaxies.

From our model, we do not expect a large difference between the
  abundance of the fully mixed  \hii{} region (reasonably well approximated
  by the abundances derived from CELs) and the general ISM. On the
  other hand, ISM abundances as measured by absorption lines are not
  likely to be significantly affected by the presence of metal-rich
  droplets.  On the other hand, the amount of O depletion of dust in
  the neutral ISM and in HII regions could be different, being
  more important in the neutral ISM than in ionized gas. Present
  determinations of the gaseous O/H in the solar vicinity from CELs in
  HII regions (Deharveng et al 2000) and from neutral absorption lines
  (Meyer et~al.\@ 1998, Jensen, Rachford, \& Snow 2005) give similar
  values.

In order to proceed further with this question of
small-scale abundance inhomogeneities, which is complicated and
involves many parameters, one needs as many observational constraints
as possible.
\begin{itemize} 
\item A systematic search for ADFs for several elements (N, Ne) in
  the same regions would be welcome. If the droplet scenario is
  valid, ratios of ADFs for several elements in the same region should
  be in agreement with the prediction of current theories of
  nucleosynthesis in massive stars. On the other hand, if
    nitrogen is mainly produced by intermediate mass stars, the
    nitrogen abundance should not be greatly enhanced in the droplets
    we consider. The measurement of nitrogen recombination lines is
    crucial to test our scenario. 
\item A systematic search for ADFs in regions of 
different metallicities would be useful.
\item Our simulations predict that values of 
ADF(\Opp) are significantly larger than one only 
in specific cases. So far,
ADFs have been measured in only a handful of 
\hii{} regions. It would be advisable to have a 
much larger number of relevant observations,
and to be able to relate the values of the ADFs 
with other properties of the \hii{} regions and 
with the observing conditions (e.g. whether
the slit encompasses a significant fraction of an 
\hii{} region  or whether the observed fluxes are heavily weighted by zones close to the ionization front).   It would  be very useful 
to have an observational census of the
cases where the ADFs are close to one. This 
implies the systematic publication of intensities 
(or upper limits) of recombination lines
in high signal-to-noise spectra.
\item   Reliable measurements of ADF(\Op) would 
also be important to constrain the models.
At present, there are a few measurements of 
ADF(\Op) in \hii{} regions, but they are highly 
uncertain since they rely on one or two weak 
lines from multiplet 1 at  $\lambda\sim7774$~\AA\ 
(the brightest multiplet expected to arise 
completely from recombination), which are 
severely affected by blends with telluric 
features.
\item  Observational diagnostics of densities and 
temperatures from recombination lines, such as 
those proposed by Tsamis et~al.\@ (2004) or Liu 
et~al.\@ (2006) should be useful to constrain the 
problem (see Garc\'\i a Rojas \& Esteban 2006 for 
a first attempt in
this direction).
\item
Another useful diagnostic is the temperature 
derived from the Balmer or Paschen jump. 
Recently, Guseva et~al.\@ (2006) have
estimated the Balmer jump temperature in about 50 
metal-poor \hii{} galaxies. They find that the 
Balmer jump temperatures are similar
to the \rOiii\ temperatures, contrary
to what is found in some nearby or more 
metal-rich  \hii{} regions (Gonz\'alez-Delgado 
et~al.\@ 2004, Garc\'\i a-Rojas et~al.\ 2004). 
The interpretation of this result in the 
framework of the droplet scenario requires a more
complicated modelling than that presented here, 
and is out of the scope of the present paper. We 
just checked by hand on a few cases that, in our 
scenario, it is possible to obtain values of the 
Balmer jump temperatures equal to (or even 
slightly higher than) the  \rOiii\ temperatures, 
depending on the fraction of \Oiiit\ and \Oiii\ 
light coming from the droplets. This question  
obviously requires further analysis. \end{itemize}

A large amount of observational  work thus 
remains to be done to improve our understanding 
of the conditions in which intimate mixing occurs 
in the ISM.

On the theoretical side, many questions remain
  unsettled. For example, no firm theoretical prediction exists
  for the droplet mass spectrum, which would allow comparison with the
  constraints on droplet sizes that we derive in Section~2.2.
Furthermore, one needs to better understand the properties of
turbulence in \hii{} regions since, in order for the droplets to
  survive long enough to explain the observed ADFs, it is vital that
  the regions not be \emph{too} turbulent.
  The question of the
  survival of the droplets in the neutral/molecular ISM is another
  question that needs addressing.
  The photoionization models
  currently rely on several free parameters that are not
  well-constrained by observations: one needs more robust estimates
of the integrated stellar yields as well as a better understanding of
the impact of massive stars on the ISM.

Given the relevance of all these issues to our understanding of the
chemical evolution of galaxies, this subject deserves to be explored
much further than we have been able to do in this paper.

\begin{acknowledgements} This work started from 
discussions between GS, GTT and MR at the 
Guillermo Haro Workshop, at INAOE Puebla, in July 
2004.
GTT acknowledges financial support from the 
Secretar\'\i{}a de Estado de Universidades e 
Investigaci\'on (Espa\~na)
through grant SAB2004-0189 and from  the 
Universit\'e de Paris~7 and the hospitality 
provided by the Instituto de Astrof\'\i{}sica de 
Andaluc\'\i{}a (IAA, CSIC) in Granada, Spain, and
also by the Observatoire de Paris Meudon, France. 
GS aknowledges the hospitality of the INAOE. WJH 
acknowledges support from DGAPA-UNAM, through 
project PAPIIT \mbox{112006-3}. We thank  Georges 
Alecian and Franck Le Petit for discussions, Bob 
Rubin and Yiannis Tsamis for their comments on a 
first version of the manuscript. Finally, we are 
very grateful to the referee as well as to Malcom Walmsley for their insightful 
comments which helped to improve the paper 
significantly.

\end{acknowledgements}

\end{document}